\documentclass{emulateapj}

\def\bc{$BC_V$}
\def\mbolsun{$M_{\rm bol,\sun}$}
\def\bcsun{$BC_{V,\sun}$}

\shortauthors{Torres}
\shorttitle{Bolometric corrections}

\begin{document}


\title{On the use of empirical bolometric corrections for stars}

\author{Guillermo Torres}

\affil{Harvard-Smithsonian Center for Astrophysics, 60 Garden St.,
Cambridge, MA 02138, USA; e-mail: gtorres@cfa.harvard.edu}

\begin{abstract} 
When making use of tabulations of empirical bolometric corrections for
stars (\bc), a commonly overlooked fact is that while the zero point
is arbitrary, the bolometric magnitude of the Sun (\mbolsun) that is
used in combination with such tables cannot be chosen arbitrarily.  It
must be consistent with the zero point of \bc\ so that the apparent
brightness of the Sun is reproduced. The latter is a measured
quantity, for which we adopt the value $V_{\sun} = -26.76 \pm 0.03$.
Inconsistent values of \mbolsun\ are listed in many of the most
popular sources of \bc.  We quantify errors that are introduced by not
paying attention to this detail. We also take the opportunity to
reprint the \bc\ coefficients of the often used polynomial fits by
\cite{Flower:96}, which were misprinted in the original publication.

\end{abstract}


\section{Introduction}
\label{sec:introduction}

Bolometric corrections are widely used in Astronomy to infer either
luminosities or absolute magnitudes of stars. Empirical corrections in
the visual band, \bc, are perhaps the most frequently needed, and
numerous tabulations exist in the literature that differ sometimes
significantly. This has been a persistent source of confusion among
users. The most common way in which these tables are employed is in
combination with the bolometric magnitude of the Sun, \mbolsun, which
is not a directly measured quantity. Many different values of
\mbolsun\ can be found in the literature as well, adding to the
confusion.  For a given table of bolometric corrections the choice of
\mbolsun\ is \emph{not} arbitrary, however, since there is an
observational constraint that must be satisfied, given by the
measurement of the visual brightness of the Sun. This fact is often
ignored, and as a result it is common to see inconsistent uses of \bc\
that can lead to errors in the luminosity of 10\% or larger, or errors
in $M_V$ of 0.1~mag or more. The primary motivation for this paper is
to call attention to this fact, to illustrate common misuses of some
of the most popular \bc\ tables, and to offer some perspective on the
problem.

One frequently used source of bolometric corrections is the tabulation
by \cite{Flower:96}, which is an update on his earlier work
\citep{Flower:77} based on a compilation of effective temperatures
($T_{\rm eff}$) and \bc\ determinations for a large number of stars.
Many authors find this source convenient because, in addition to the
table, it presents simple polynomial fits for \bc\ valid over a wide
range of temperatures.  Unfortunately the original publication had
misprints in the coefficients of those formulae that have prevented
their use, and an erratum was never published. The present author has
received numerous inquiries about this problem over the years. Thus, a
second motivation for this paper is to make the correct coefficients
available to the community, as well as to amend the form of the
equation and one of the coefficients for a color/temperature
calibration in the same paper that were also misprinted. We begin by
presenting these corrections, and follow with a discussion of the
practical use of \bc\ tables.

\section{Coefficients for the Flower (1996) \bc\ and $\log T_{\rm eff}$ formulae}
\label{sec:flower}

\cite{Flower:96} expressed the bolometric corrections as a function of
the effective temperature of the star, and presented polynomial fits
for \bc\ of the form
\begin{equation}
BC_V = a + b\thinspace(\log T_{\rm eff}) + c\thinspace(\log T_{\rm eff})^2 + \cdots~,
\end{equation}
for three different temperature ranges. The coefficients $a, b, c,
\cdots$ were given in his Table~6 for each interval, but are missing
powers of ten. We rectify this situation here and present them with a
larger number of significant digits (see Table~\ref{tab:bc}).  While
this formulation is certainly very handy for many applications, it
must be kept in mind that bolometric corrections become less reliable
for cooler stars due to the limited number of observational
constraints, and they break down completely for the M dwarfs.  Two of
the very few such constraints available are for the low-mass eclipsing
binaries CU~Cnc (components with $M \approx 0.42~M_{\sun}$) and CM~Dra
($M \approx 0.24~M_{\sun}$), which have known parallaxes. For these
stars one may compute the \bc\ values for the components directly from
the estimated $M_V$ values and the bolometric magnitudes based on the
absolute radius and temperature determinations \citep{Ribas:03,
Morales:09}. Comparing the measured values with the predicted ones
shows that the Flower relations give a \bc\ that is too negative by
$\sim$1.2 mag for CU~Cnc, and $\sim$1.7 mag for CM~Dra. Similar
discrepancies are found in other tabulations of empirical bolometric
corrections.

\begin{deluxetable*}{cccc}
\tabletypesize{\scriptsize}
\tablewidth{0pc}

\tablecaption{Bolometric corrections by \cite{Flower:96} as a function of temperature:
$BC_V = a + b(\log T_{\rm eff}) + c(\log T_{\rm eff})^2 +
\cdots$\label{tab:bc}}

\tablehead{
\colhead{Coefficient} & 
\colhead{$\log T_{\rm eff} < 3.70$} &
\colhead{$3.70 < \log T_{\rm eff} < 3.90$} & 
\colhead{$\log T_{\rm eff} > 3.90$} 
}
\startdata
 $a$  &  $-$0.190537291496456E+05   &  $-$0.370510203809015E+05   &  $-$0.118115450538963E+06    \\
 $b$  &  \phs0.155144866764412E+05  &  \phs0.385672629965804E+05  &  \phs0.137145973583929E+06	\\
 $c$  &  $-$0.421278819301717E+04   &  $-$0.150651486316025E+05   &  $-$0.636233812100225E+05	\\
 $d$  &  \phs0.381476328422343E+03  &  \phs0.261724637119416E+04  &  \phs0.147412923562646E+05	\\
 $e$  &  \nodata   		    &  $-$0.170623810323864E+03   &  $-$0.170587278406872E+04	\\
 $f$  &   \nodata                   &  \nodata                    &  \phs0.788731721804990E+02  \\ [-1.5ex]
\enddata
\end{deluxetable*}

A useful color/temperature calibration was also presented by
\cite{Flower:96} in his Table~5, separately for supergiants and for
main-sequence, subgiant, and giant stars, but unfortunately the
formula given there was printed incorrectly, and should have expressed
the temperature ($\log T_{\rm eff}$) as a function of the $B\!-\!V$
color index, rather than the reverse. The polynomial fits are similar
to those used for \bc:
\begin{equation}
\log T_{\rm eff} = a + b\thinspace(B\!-\!V) + c\thinspace(B\!-\!V)^2 + \cdots~.
\end{equation}
One of original coefficients was also misprinted, and a correction was
issued by \cite{Prsa:05}. We present all coefficients again in
Table~\ref{tab:bmv} to higher precision.

\begin{deluxetable}{ccc}
\tablewidth{0pc}

\tablecaption{Effective temperature as a function of color \citep{Flower:96}: $\log
T_{\rm eff} = a + b(B\!-\!V) + c(B\!-\!V)^2 + \cdots$\label{tab:bmv}}

\tablehead{
\colhead{} & 
\colhead{} & 
\colhead{Main-Sequence Stars,} \\
\colhead{Coefficient} & 
\colhead{Supergiants} & 
\colhead{Subgiants, Giants}
}
\startdata
 $a$  &     \phs4.012559732366214   &   \phs3.979145106714099 \\
 $b$  &     $-$1.055043117465989   &   $-$0.654992268598245 \\
 $c$  &     \phs2.133394538571825   &    \phs1.740690042385095 \\
 $d$  &     $-$2.459769794654992   &   $-$4.608815154057166 \\
 $e$  &     \phs1.349423943497744   &    \phs6.792599779944473 \\
 $f$  &     $-$0.283942579112032   &   $-$5.396909891322525 \\
 $g$  &     \nodata              &    \phs2.192970376522490 \\
 $h$  &     \nodata              &   $-$0.359495739295671 \\ [-1.5ex]
\enddata
\end{deluxetable}
\vskip 1.5cm

\section{The use of empirical bolometric corrections}
\label{sec:use}

As is well known, the apparent bolometric magnitude of a star is
defined as
\begin{equation}
\label{eq:mbol}
m_{\rm bol} = -2.5 \log\left(\int_{0}^{\infty}f_{\lambda} d\lambda\right) + C_1~,
\end{equation}
where $f_\lambda$ is the monochromatic flux from the object per unit
wavelength interval received outside the Earth's atmosphere, and $C_1$
is a constant. The bolometric correction is usually defined as the
quantity to be added to the apparent magnitude in a specific passband
(in the absence of interstellar extinction) in order to account for
the flux outside that band:
\begin{equation}
\label{eq:bc}
BC_V =  m_{\rm bol} - V  =  M_{\rm bol} - M_V~.
\end{equation}
We focus here on the visual band because that is the context in which
bolometric corrections were historically defined, although of course
the definition can be generalized to any passband. Note that this
definition is usually interpreted to imply that the bolometric
corrections must always be negative, although many of the currently
used tables of empirical \bc\ values violate this condition.  We
return to this below.  Eq.(\ref{eq:bc}) may also be written as
\begin{equation}
BC_V = 2.5 \log\left(\textstyle \int_{0}^{\infty}S_{\lambda}(V) f_{\lambda} d\lambda~/ \textstyle\int_{0}^{\infty} f_{\lambda} d\lambda \right) + C_2~,
\end{equation}
where $S_{\lambda}(V)$ is the sensitivity function of the $V$
magnitude system. The constant $C_2$ contains an arbitrary zero point
that has been a common source of confusion. This zero point has
traditionally been set using the Sun as the reference. By noting that
\begin{equation}
\label{eq:bcsun}
BC_{V,\sun} =  m_{\rm bol,\sun} - V_{\sun} = M_{\rm bol,\sun} - M_{V,\sun}~,
\end{equation}
it is immediately clear that setting a value for the bolometric
correction of the Sun is equivalent to specifying the zero point of
the bolometric magnitude scale, since $V_{\sun}$ is a known
quantity. A common practice when using one of the many available
tables of empirical bolometric corrections is to \emph{adopt} a value
for \mbolsun, but all too often this is done without regard for
whether the chosen value is consistent with \bcsun\ \emph{from the
same table}. From eq.(\ref{eq:bcsun}) we have
\begin{equation}
\label{eq:mbolsun}
M_{\rm bol,\sun} = M_{V,\sun} + BC_{V,\sun} = V_{\sun} + 31.572 + BC_{V,\sun}~,
\end{equation}
where the numerical constant (with the opposite sign) corresponds to
the distance modulus of the Sun at 1~AU.  This formulation shows that
once a particular tabulation of $BC_V$ values is adopted, the absolute
bolometric magnitude of the Sun is no longer arbitrary. This fact has
been emphasized by \cite{Bessell:98}, and other authors, but is still
largely overlooked.

Direct measurements of $V_{\sun}$ are difficult to make because of the
extreme difference in brightness between the Sun and the stars that
are used as the reference in the $V$ system, and also because the Sun
is spatially resolved. Nevertheless, many careful determinations have
been carried out over the years (although not very recently, as far as
we are aware), and the most reliable of the photoelectric measurements
have been reviewed by \cite{Hayes:85}.\footnote{A useful compilation
and discussion of solar data may be found in ``Basic Astronomical Data
for the Sun (BADS?)'', maintained by Eric Mamajek, Univ.\ of Rochester
(NY), at \tt{http://www.pas.rochester.edu/$\sim$emamajek/sun.txt}.}
Among them, one that carries particularly high weight is that of
\cite{Stebbins:57}, which was adjusted slightly (by $-0.02$~mag) by
\cite{Hayes:85} for an error in the treatment of horizontal
extinction, and further updated by \cite{Bessell:98} using modern $V$
magnitudes for the reference stars. The result is $V_{\sun} = -26.76
\pm 0.03$.\footnote{The uncertainty we assign is slightly larger than
the $\pm0.02$~mag given by \cite{Bessell:98} because it includes
contributions from systematic effects described by
\cite{Stebbins:57}. We also re-examined the corrections made in the
later work to reduce the measurements to the distance of 1~AU, since
variations in the Earth-Sun distance throughout the year lead to
non-negligible brightness changes of $\pm0.036$~mag. This leads to
only a very minor difference in the third decimal place compared to
the value reported by \cite{Bessell:98}, which we have ignored here.}
Two additional determinations discussed by \cite{Hayes:85} are those
of \cite{Nikonova:49}, transformed to the standard Johnson system by
\cite{Martynov:60}, and \cite{Gallouet:64}: $V_{\sun} = -26.81 \pm
0.05$ and $V_{\sun} = -26.70 \pm 0.01$, respectively. Because
systematic effects likely dominate the differences, we follow
\cite{Hayes:85} and adopt as a consensus value a simple average of the
three estimates, giving $V_{\sun} = -26.76 \pm 0.03$, with an error
that is probably realistic. The absolute visual magnitude of the Sun
then becomes $M_{V,\sun} = 4.81 \pm 0.03$.

Measurements of $V_{\sun}$ based on absolute flux-calibrated spectra
of the Sun \citep[][and others]{Colina:96, Thuillier:04} or synthetic
spectra based on model atmospheres such as ATLAS9 and MARCS have also
been made by many authors, and generally range from $-26.74$ to
$-26.77$, with minor differences depending on the author even when
using the same spectrum \citep[see, e.g.,][]{Bessell:98,
Casagrande:06}.  These estimates agree well with the direct
measurements.

While the zero point of \bc\ is completely arbitrary, and no
particular scale has been officially endorsed by the International
Astronomical Union (IAU) (but see Sect.~\ref{sec:discussion}), it is
common for some authors of these tabulations to define the scale by
\emph{adopting} a certain value for \bcsun, sometimes for historical
reasons. For example, the widely used reference by \cite{Cox:00}, the
successor to Allen's Astrophysical Quantities \citep[][and earlier
editions]{Allen:76} indicates that it adopts $BC_{V,\sun} = -0.08$
(p.\ 341, but see also the next section), a value inherited from
earlier compilations. Similar scales have been chosen in many other
empirical tables. On the other hand, the theoretical \bc\ values that
are incorporated in the stellar evolution models of
\cite{Pietrinferni:04} are based on $BC_{V,\sun} = -0.203$, while the
\cite{Yi:01} isochrones use either $BC_{V,\sun} = -0.109$ or
$BC_{V,\sun} = -0.08$, for the two different color transformation
tables offered with their models \citep{Lejeune:98, Green:87}. The
more negative values above tend to come from the practice by many
stellar modelers \citep[starting with][if not earlier]{Buser:78} of
computing theoretical bolometric corrections from a grid of model
atmospheres for a large range of metallicities, temperatures, and
surface gravities, and then shifting all values by arbitrarily setting
the smallest \bc\ to zero. This usually corresponds to late A-type
stars of low surface gravity.\footnote{An example of a table of
theoretical bolometric corrections often used is that of
\cite{Lejeune:98}, mentioned above. These authors initially followed
the procedure just described, leading to $BC_{V,\sun} = -0.190$, but
then chose to adjust the zero point to provide the best fit to the
\bc\ table of \cite{Flower:96}.}

The scale adopted by \cite{Flower:96} is such that $BC_{V,\sun} =
-0.080$. As a result, his bolometric corrections for stars between
$T_{\rm eff} \approx 6400$~K and $T_{\rm eff} \approx 8500$~K
(spectral type approximately F5 to A5) are \emph{positive}, seemingly
conflicting with the idea that bolometric magnitudes ought to be
brighter than $V$ magnitudes, according to eq.(\ref{eq:bc}).  Other
tables with a similar zero point as Flower's share the same
problem. In reality, however, the contradiction is of no consequence
because of the arbitrary nature of the zero point. Luminosities
inferred for stars are never affected if a consistent value of
\mbolsun\ is used, because the bolometric magnitudes are always
compared to the Sun.  To see this more clearly one may make use of
\begin{equation}
M_{\rm bol} - M_{\rm bol,\sun} = -2.5 \log(L/L_{\sun})
\end{equation}
along with eq.(\ref{eq:bc}) and eq.(\ref{eq:mbolsun}) to express the
luminosity of a star in terms of its absolute visual magnitude and
bolometric correction as
\begin{equation}
\label{eq:lum}
\log(L/L_{\sun}) = -0.4\left[ M_V - V_{\sun} - 31.572 + (BC_V - BC_{V,\sun})\right]~.
\end{equation}
Alternately, if seeking to determine the absolute magnitude from
knowledge of the luminosity (e.g., in eclipsing binaries if the
temperature and absolute radius are known), one has
\begin{equation}
\label{eq:mv}
M_V = -2.5 \log(L/L_{\sun}) + V_{\sun} + 31.572 - (BC_V - BC_{V,\sun})~.
\end{equation}
The last two equations involve only the \emph{difference} between two
bolometric corrections, so that the zero point cancels out. The
apparent contradiction is thus irrelevant since any table of \bc\ may
be shifted arbitrarily with no impact on the results from these
expressions.

\section{Inconsistencies in published tables of empirical \bc}
\label{sec:published}

As an illustration of the confusing state of affairs brought about by
the proliferation of zero points, and to make readers aware of the
sometimes serious inconsistencies lurking in these \bc\ sources, we
examine here several of the widely used tables of empirical bolometric
corrections and spell out their assumptions.  It is not uncommon for
tabular information of this kind to be copied over from earlier
sources, but assumptions are sometimes changed along the way, so that
each table has its own problems:

\noindent$\bullet$~In the chapter on the Sun, the latest edition of
the popular Allen's Astrophysical Quantities \citep[][p.\ 341]{Cox:00}
adopts an internally consistent set of solar parameters given by
$BC_{V,\sun} = -0.08$, $M_{\rm bol,\sun} = 4.74$, $M_{V,\sun} = 4.82$,
and $V_{\sun} = -26.75$.  However, inspection of the table of
bolometric corrections for dwarfs in the chapter on normal stars (p.\
388), which is the one employed in practice, reveals that the \bcsun\
there is $-0.20$ rather than the value advocated earlier. This implies
$V_{\sun} = -26.63$. Therefore, the use of this table together with
$M_{\rm bol,\sun} = 4.74$ will introduce a systematic error of
0.13~mag, which is the same as produced by a difference of 500~K in
the input temperature. In order to be consistent with the measured
value of $V_{\sun} = -26.76$ discussed in the previous section, the
bolometric magnitude that should be used for the Sun is $M_{\rm
bol,\sun} = 4.61$. An earlier edition of Allen's Astrophysical
Quantities has inconsistencies of its own and adopts a rather
different \bc\ scale which, interestingly, does not have as serious a
problem. For example, the solar parameters in the 3rd edition by
\cite{Allen:76} are also internally consistent, and again use
$BC_{V,\sun} = -0.08$, but with $M_{\rm bol,\sun} = 4.75$ (p.\ 162 of
that work). The \bc\ table on p.\ 206, however, lists a bolometric
correction for a normal star with the solar temperature as
$BC_{V,\sun} = -0.05$. Nevertheless, if used in conjunction with the
solar bolometric magnitude advocated, this table implies $V_{\sun} =
-26.77$, which is much more accurate than in the later edition.

\noindent$\bullet$~The solar values adopted by \cite{Schmidt-Kaler:82}
(p.\ 451) are $BC_{V,\sun} = -0.19$, $M_{\rm bol,\sun} = 4.64$,
$M_{V,\sun} = 4.83$ and $V_{\sun} = -26.74$, which are internally
consistent. Their \bc\ table for main-sequence stars on p.\ 453 gives
a slightly different value of $BC_{V,\sun} = -0.21$ for a star of
solar temperature. This implies $V_{\sun} = -26.72$ rather than their
adopted value. To be consistent with $V_{\sun}$ from the previous
section, the bolometric magnitude to be used for the Sun is $M_{\rm
bol,\sun} = 4.60$.

\begin{deluxetable*}{lccccc}[!t]
\tablewidth{0pc}

\tablecaption{Empirical \bc\ scales and \mbolsun\ values from the
literature\label{tab:bcsum}}

\tablehead{
\colhead{} & 
\colhead{Advocated} & 
\colhead{Actual} & 
\colhead{Adopted} & 
\colhead{Recommended} & 
\colhead{} \\
\colhead{} & 
\colhead{$BC_{V,\sun}$} & 
\colhead{$BC_{V,\sun}$} & 
\colhead{\mbolsun} & 
\colhead{\mbolsun} & 
\colhead{Error} \\
\colhead{Source} & 
\colhead{(mag) \tablenotemark{a}} & 
\colhead{(mag) \tablenotemark{b}} & 
\colhead{(mag) \tablenotemark{c}} & 
\colhead{(mag) \tablenotemark{d}} & 
\colhead{(mag) \tablenotemark{e}}
}
\startdata
\cite{Cox:00}           & $-$0.08 & $-$0.20 & 4.74    & 4.61 & $+$0.13   \\
\cite{Allen:76}         & $-$0.08 & $-$0.05 & 4.75    & 4.76 & $-$0.01   \\
\cite{Schmidt-Kaler:82} & $-$0.19 & $-$0.21 & 4.64    & 4.60 & $+$0.04   \\
\cite{Lang:92}          & $-$0.07 & $-$0.20 & 4.75    & 4.61 & $+$0.14   \\
\cite{Popper:80}        & $-$0.14 & $-$0.14 & 4.69    & 4.67 & $+$0.02   \\
\cite{Gray:05}          & \nodata & $-$0.09 & 4.73    & 4.72 & $+$0.01   \\
\cite{Straizys:80}      & \nodata & $-$0.07 & 4.72    & 4.74 & $-$0.02   \\
\cite{Kenyon:95}        & \nodata & $-$0.21 & \nodata & 4.60 & \nodata   \\
\cite{Flower:96}        & \nodata & $-$0.08 & \nodata & 4.73 & \nodata   \\ [-1.5ex]
\enddata
\tablenotetext{a}{Value that each source states to have adopted as the zero point of their \bc\ scale.}
\tablenotetext{b}{Value read off from the relevant \bc\ table for each source.}
\tablenotetext{c}{Bolometric correction for the Sun adopted by each source.}
\tablenotetext{d}{\mbolsun\ value required for consistency with $V_{\sun} = -26.76$ (Sect.~\ref{sec:use}), when using the \bc\ table as published.}
\tablenotetext{e}{Error incurred when using the published \bc\ table combined with \mbolsun\ from the source, instead of the recommended \mbolsun\ value in the previous column.}
\end{deluxetable*}

\noindent$\bullet$~The extensive compilation by \cite{Lang:92} adopts
the following values for the Sun (p.\ 103): $V_{\sun} = -26.78$,
$M_{V,\sun} = 4.82$, and $M_{\rm bol,\sun} = 4.75$. The first two are
slightly inconsistent with the distance modulus of the Sun. The last
two imply $BC_{V,\sun} = -0.07$, yet the listing of bolometric
corrections on p.\ 138 gives $BC_{V,\sun} = -0.20$ for a star of solar
temperature. According to eq.(\ref{eq:mbolsun}), the proper value of
\mbolsun\ to use with this table is $M_{\rm bol,\sun} =
4.61$. Consequently, the systematic error incurred by using the
\cite{Lang:92} table in combination with their \mbolsun\ is 0.14~mag.

\noindent$\bullet$~A \bc\ table still often used mainly in the binary
star field, is that of \cite{Popper:80}. The bolometric correction and
absolute visual magnitude adopted there for the Sun are $BC_{V,\sun} =
-0.14$ and $M_{V,\sun} = 4.83$, from which $V_{\sun} = -26.74$. These
imply $M_{\rm bol,\sun} = 4.69$. To be in exact agreement with our
apparent magnitude for the Sun, the solar bolometric magnitude should
be adjusted slightly to $M_{\rm bol,\sun} = 4.67$.

\noindent$\bullet$~The \bc\ table in the textbook by \cite{Gray:05}
gives $BC_{V,\sun} = -0.09$ for a star of solar temperature, and
adopts $V_{\sun} = -26.75$ (and the corresponding value of $M_{V,\sun}
= 4.82$). Together these imply $M_{\rm bol,\sun} = 4.73$. For exact
consistency with $V_{\sun}$ from the previous section, we recommend
using $M_{\rm bol,\sun} = 4.72$.

\noindent$\bullet$~The work of \cite{Straizys:80} contains many useful
tables of average stellar properties, and adopts a zero point for the
\bc\ scale that is adjusted to give $BC_{V,\sun} = -0.07$. These
authors also adopt $M_{\rm bol,\sun} = 4.72$, which leads to
$M_{V,\sun} = 4.79$ and $V_{\sun} = -26.78$. Perfect agreement with
our $V_{\sun}$ requires $M_{\rm bol,\sun} = 4.74$.

\noindent$\bullet$~\cite{Kenyon:95} presented a table of bolometric
corrections that is often used in the field of pre-main sequence
stars, and has been incorporated into some model isochrones for young
stars such as those by \cite{Siess:00}. It is compiled from a variety
of sources, and the zero point is such that a star of solar
temperature has $BC_{V,\sun} = -0.21$. No value of \mbolsun\ is
specified.

\noindent$\bullet$~Finally, as mentioned earlier, the \bc\ table by
\cite{Flower:96} gives $BC_{V,\sun} = -0.08$ for a star of solar
temperature, but there is no associated value of \mbolsun\ given in
the text.  For consistency with $V_{\sun}$ from the previous section,
the number to be used is $M_{\rm bol,\sun} = 4.73$.

Table~\ref{tab:bcsum} summarizes the various empirical \bc\ scales,
along with the \mbolsun\ values listed by each source, as well as the
\mbolsun\ recommended here to maintain consistency with the adopted
$V_{\sun}$. The systematic error introduced when using the former
instead of the latter is presented in the last column.

\section{Discussion}
\label{sec:discussion}

From a practical point of view one may approach the use of published
tables of empirical \bc\ values in several ways, but it is essential
to always maintain consistency with the measured brightness of the
sun, $V_{\sun}$. Errors introduced when overlooking this requirement
are seen rather often in the literature, and are quantified in
Table~\ref{tab:bcsum}.  \cite{Bessell:98} chose to \emph{define}
$M_{\rm bol,\sun} = 4.74$, from which one derives $BC_{V,\sun} =
-0.07$. If one follows this path then it is necessary to adjust the
\bc\ table one is using, in order to match this zero point. In general
each table will require a different offset. However, most users find
it more convenient to adopt a particular \bc\ table `as is', in which
case care must be taken to use the proper \mbolsun\ as listed in
Table~\ref{tab:bcsum}, and not an arbitrary value from another source
(or even from the same source if it is inconsistent).  Yet another
approach is to use eq.(\ref{eq:lum}) or eq.(\ref{eq:mv}) directly, as
advocated by \cite{Gray:05}, which dispenses with having to find a
formal \mbolsun\ value, and requires only to read off \bc\ for the Sun
from the same table used for the star of interest.  These three
approaches are of course completely equivalent.

Somewhat surprisingly the IAU has not issued a formal resolution on
the matter of \bc\ zero points, although two of its Commissions did
agree at the Kyoto meeting of 1997 \citep[][pp.\ 141 and
181]{Andersen:99} on a preferred scale that is equivalent to adopting
a value for \mbolsun. The scale was set by \emph{defining} a star with
$M_{\rm bol} = 0.00$ to have an absolute radiative luminosity of $L =
3.055 \times 10^{28}$ W \citep[see also][]{Cayrel:02}. The rationale
was that this value together with the nominal bolometric luminosity of
the Sun adopted by the international GONG project ($L_{\sun} = 3.846
\times 10^{26}$ W, according to the IAU Commission reports cited
above) leads exactly to $M_{\rm bol,\sun} = 4.75$, which is the
bolometric magnitude for the Sun listed in the 1976 edition of
Astrophysical Quantities \citep{Allen:76}. This was a widely used
source at the time (and still is, by some), so it was thought to be a
logical choice. Effectively, therefore, the scale is set by this value
of \mbolsun. Combined with our adopted solar brightness of $V_{\sun} =
-26.76$, it implies $BC_{V,\sun} = -0.06$.  As it turns out, however,
the most recent edition of Astrophysical Quantities \citep{Cox:00} did
not follow that recommendation, and adopted a slightly different zero
point. So have some other recent \bc\ compilations (see
Table~\ref{tab:bcsum}).


In practice the adoption of a value of \bcsun\ or a value of \mbolsun\
may not be the most convenient way to solve the immediate problem
faced by users.  The first alternative would not be of much help to
those wishing to make use of an existing \bc\ table, and the second
would force them to adjust the table to match $V_{\sun}$. To conclude,
one may argue that it would perhaps be more useful instead to agree on
the best value for the apparent visual magnitude of the Sun, which is
directly measured. 


\acknowledgements

I am indebted to Phillip Flower for providing me with the correct
coefficients for his \bc\ and color/temperature relations, which were
misprinted in his original work \citep{Flower:96}. They are presented
here with his permission. I also thank Gene Milone for stimulating
discussions and motivation for this paper, Todd Henry for alerting me
to the significant errors in \bc\ for cool stars, and the anonymous
referee for helpful comments. Correspondence with Martin Asplund,
Dainis Dravins, Arlo Landolt, Pedro Mart\'\i nez, Gene Milone, and
Chris Sterken regarding IAU deliberations on the matter of \bc\ is
also acknowledged. This work was partially supported by NSF grant
AST-0708229. The research has made use of NASA's Astrophysics Data
System Abstract Service.







\begin{thebibliography}{}

\bibitem[Allen(1976)]{Allen:76}
 Allen, C.\ W. 1976, Astrophysical Quantities, 3rd Ed. (London: The
 Athlone Press)

\bibitem[Andersen(1999)]{Andersen:99}
 Andersen, J. 1999, Proceedings of the Twenty-Third General Assembly,
 Transactions of the IAU, Vol.\ XXIIIB (Dordrecht: Kluwer)



\bibitem[Bessell et al.(1998)]{Bessell:98}
 Bessell, M.\ S., Castelli, F., \& Plez, B. 1998, \aap, 333, 231

\bibitem[Buser \& Kurucz(1978)]{Buser:78}
 Buser, R., \& Kurucz, R.\ L. 1978, \aap, 70,555

\bibitem[Casagrande et al.(2006)]{Casagrande:06}
 Casagrande, L., Portinari, L., \& Flynn, C. 2006, \mnras, 373, 13

\bibitem[Cayrel(2002)]{Cayrel:02}
 Cayrel, R. 2002, in Observed HR diagrams and stellar evolution, ASP
 Conf.\ Ser.\ 274, eds.\ T. Lejeune \& J. Fernandes (San Francisco:
 ASP), p.\ 133

\bibitem[Colina et al.(1996)]{Colina:96}
 Colina, L., Bohlin, R.\ C., \& Castelli, F. 1996, \aj, 112, 307

\bibitem[Cox(2000)]{Cox:00} 
 Cox, A.\ N. 2000, Allen's Astrophysical Quantities, 4th Ed. (Berlin:
 Springer)

\bibitem[Flower(1977)]{Flower:77}
 Flower, P.\ J. 1977, \aap, 54, 31

\bibitem[Flower(1996)]{Flower:96}
 Flower, P.\ J. 1996, \apj, 469, 355

\bibitem[Gallou\"et(1964)]{Gallouet:64}
 Gallou\"et, L. 1964, An.\ Ap., 27, 423

\bibitem[Gray(2005)]{Gray:05}
 Gray, D.\ F. 2005, The Observation and Analysis of Stellar
 Photospheres (Cambridge: CUP), p.\ 506

\bibitem[Green et al.(1987)]{Green:87}
 Green, E.\ M., Demarque, P., \& King, C.\ R. 1987, The Revised Yale
 Isochrones and luminosity Functions (New Haven: Yale Univ.\ Obs.)

\bibitem[Hayes(1985)]{Hayes:85} Hayes, D.\ S. 1985, in IAU Symp.\ 111,
 Calibration of Fundamental Stellar Quantities, eds.\ D.\ S.\ Hayes et
 al.\ (Reidel: Dordrecht), p.\ 225

\bibitem[Kenyon \& Hartmann(1995)]{Kenyon:95}
 Kenyon, S.\ J., \& Hartmann, L. 1995, \apjs, 101, 117

\bibitem[Lang(1992)]{Lang:92}
 Lang, K.\ R. 1992, Astrophysical Data: Planets and Stars (New York:
 Springer-Verlag)

\bibitem[Lejeune et al.(1998)]{Lejeune:98}
 Lejeune, Th., Cuisinier, F., \& Buser, R. 1998, \aap, 130, 65

\bibitem[Martynov(1960)]{Martynov:60}
 Martynov, D.\ Ya. 1960, \sovast, 3, 633

\bibitem[Morales et al.(2009)]{Morales:09}
 Morales, J.\ C., Ribas, I., Jordi, C., Torres, G., Gallardo, J.,
Guinan, E.\ F., Charbonneau, D., Wolf, M., Latham, D.\ W.,
Anglada-Escud\'e, G., Bradstreet, D.\ H., Everett, M.\ E., O'Donovan,
F.\ T., Mandushev, G., \& Mathieu, R.\ D. 2009, ApJ, 691, 1400

\bibitem[Nikonova(1949)]{Nikonova:49}
 Nikonova, E.\ K. 1949, Izv.\ Krymsk.\ Astrofiz.\ Observ., 4, 114

\bibitem[Pietrinferni et al.(2004)]{Pietrinferni:04}
 Pietrinferni, A., Cassisi, S., Salaris, M., \& Castelli, F. 2004,
 \apj, 642, 797

\bibitem[Popper(1980)]{Popper:80}
 Popper, D.\ M. 1980, \araa, 18, 115

\bibitem[Pr\v sa \& Zwitter(2005)]{Prsa:05}
 Pr\v sa, A., \& Zwitter, T. 2005, \apj, 628, 426

\bibitem[Ribas(2003)]{Ribas:03}
 Ribas, I. 2003, \aap, 398, 239

\bibitem[Schmidt-Kaler(1982)]{Schmidt-Kaler:82} 
 Schmidt-Kaler, Th. 1982, in Landolt-B\"ornstein, Numerical Data and
 Functional Relationships in Science and Technology, Vol.\ 2, eds.\
 K. Schaifers \& H.\ H.\ Voigt (Berlin: Springer)

\bibitem[Siess et al.(2000)]{Siess:00}
 Siess, L., Dufour, E., \& Forestini, M. 2000, \aap, 358, 593

\bibitem[Stebbins \& Kron(1957)]{Stebbins:57}
 Stebbins, J., \& Kron, G.\ E. 1957, \apj, 126, 266

\bibitem[Strai\v zys \& Kuriliene(1980)]{Straizys:80}
 Strai\v zys, V., \& Kuriliene, G. 1980, \apss, 80, 353

\bibitem[Thuillier et al.(2004)]{Thuillier:04}
 Thuillier, G., Floyd, L., Woods, T.\ N., Cebula, R., Hilsenrath, E.,
 Hers\'e, M., \& Labs, D. 2004, Adv.\ Space Res., 34, 256



\bibitem[Yi et al.(2001)]{Yi:01}
 Yi, S.\ K., Demarque, P., Kim, Y.-C., Lee, Y.-W., Ree, C.\ H.,
Lejeune, T., \& Barnes, S. 2001, \apjs, 136, 417


\end{thebibliography}
\end{document}